# Evidence for radio loud to radio quiet evolution from red and blue quasars

David Garofalo & Katie Bishop

Department of Physics, Kennesaw State University

Abstract

Recent work on red and blue quasi-stellar objects (QSOs) has identified peculiar number distributions as a function of radio loudness that we explore and attempt to explain from the perspective of a picture in which a subset of the population of active galaxies evolves from the radio loud to the radio quiet state. Because the time evolution is slowed down by an order of magnitude or more for the radio quiet phase, the numbers of red and blue QSOs approach each other at the extreme end of the radio quiet range of radio loudness with larger numbers. The rapid time evolution of most radio loud phases, instead, makes the numbers similar but lower at the far radio loud end. At the midpoint of radio loudness, instead, the differences between red and blue QSOs experience their largest values which results from accretion rapidly spinning black holes down but subsequently spinning them up more slowly. Recovering these basic features of the observations is evidence for rapid evolution away from radio loudness and slow evolution in radio quiet states.

1. Introduction

Rosario et al (2020) use LOFAR and Fawcett et al (2020) use VLA data from the FIRST, Stripe82, and COSMOS surveys to explore differences in red and blue quasars (QSOs), concluding that red QSOs have both a higher radio detection rate and incidence of compact radio morphology, while the differences disappear at the extremes of radio loudness. Compared to blue QSOs, red QSOs appear to be merger-triggered (Urrutia, Lacy & Becker 2008; Glikman et al 2015), and to be about 10% of the total (Richards et al 2003; Klindt et al 2019; while Lacy et al 2013 claim more than 20%), with more powerful outflows (Urrutia 2009). The idea that mergers trigger active galactic nuclei (AGN) has a long history (e.g. Barnes & Hernquist 1991). A consequence of this idea is that the most powerful AGN, known as quasars, should be predominantly found in merger prone environments, i.e. dense groups or low mass clusters. This is not born out in the data (e.g. Wylezalek et al 2013; Zakamska et al 2019).

In this work we show how the results of Rosario et al (2020) and Fawcett et al (2020) support a picture in which both the most radio loud quasars or QSOs and the most radio quiet quasars or QSOs result from accretion onto rapidly rotating black holes in retrograde and prograde configurations, respectively. Intrinsic to this picture is the fact that retrograde accretion spins black holes toward prograde accretion, implying that radio loud quasars, if formed, constitute initial states, but never final ones. If the red quasars, or rQSOs, of Rosario et al (2020) and Fawcett et al (2020) are quasars triggered in mergers, they will have a range of black hole spin values across the retrograde to prograde spectrum. Hence, a range of radio loud and radio quiet quasars. This distribution, however, can only evolve over time through the accretion process which spins retrograde black holes down and eventually up into the prograde regime. As a consequence of this simple time evolution, the rQSO distribution in radio loudness must evolve in a way that reflects the change in black hole spin values over time. For accreting black holes that remain quasar-like over time, the distribution is easy to predict and we describe that evolution over 10 million to 100 million years. Our claim is that these later time distributions are the cQSOs, or blue quasars, of Rosario et al (2020). In section 2 we illustrate the observations and the model designed to explain them and in Section 3 we conclude.

2. Discussion

In Figure 1 we reproduce the ratio of the numbers of rQSOs and cQSOs of Rosario et al (2020) as a function of radio loudness. As pointed out in their work, the ratio is significantly removed from value unity toward the radio loud/radio quiet boundary or midpoint of Figure 1 shown with a dashed line. The values of Figure 1 come from the numbers of rQSOs and cQSOs from Rosario et al (2020) and which we separately show in Figure 2. Because the numbers reflect 6 micron, redshift matched, LOFAR detections, the actual distributions are enhanced with respect to Figure 2. And since radio quiet sources are in the majority, the actual distribution of cQSOs should show a greater enhancement and therefore likely exceed that for rQSOs regardless of radio loudness. We will show how to understand that feature as well as the tendency of the rQSOs to cQSOs ratio to peak in the radio loud range as shown in Figure 1.

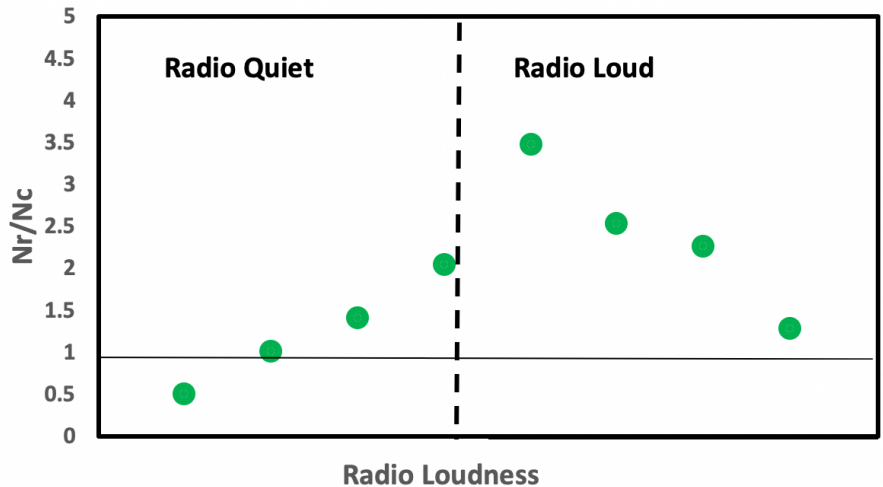

Figure 1: Ratio of LoTSS detection rate for rQSOs and cQSOs versus radio loudness from Rosario et al (2020). While the numbers approach unity at the extremes of radio loudness, they differ from that significantly in the middle.

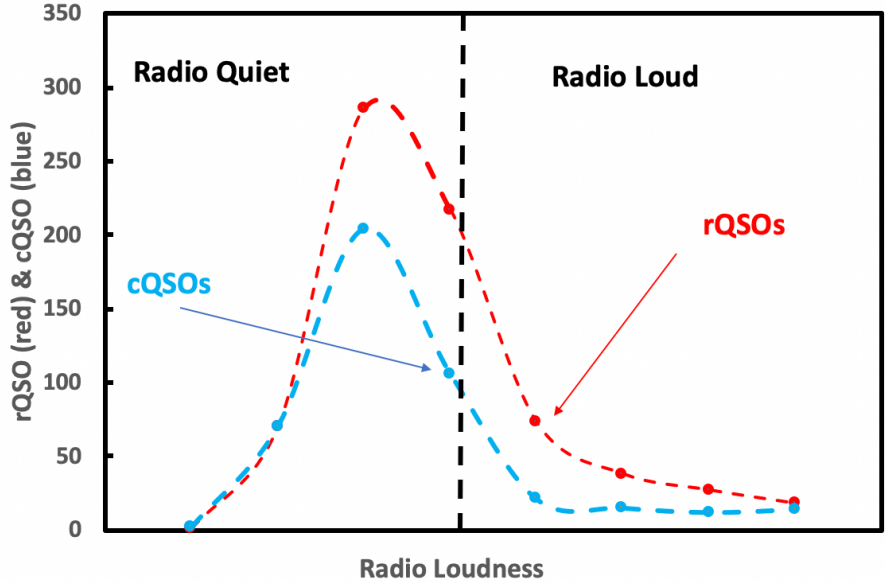

Figure 2: Numbers of rQSOs (red) and cQSOs (blue) as a function of radio loudness reported in Rosario et al (2020), the ratio of which is plotted in Figure 1. Data points are smoothly connected for ease of comparison.

    We will now show how to construct the red and blue curves that make sense of Figure 1 from a model based on the idea that greatest radio loudness emerges from merger-triggered nuclei with cold gas funneled onto high spinning black holes in retrograde configuration (Garofalo, Evans & Sambruna 2010). Among the many recent applications of this model, there has been a resolution to the aforementioned absence of direct evidence for merger signatures associated with both radio loud and radio quiet quasars (Garofalo, Webster & Bishop 2020). Figure 3 shows the characteristic time evolution of originally retrograde accreting black holes following mergers in less rich environments, one of the possible paths for retrograde-triggered black holes that apply to the rQSOs of Rosario et al (2020). Because of abundant cold gas, the initial state is radiatively efficient which merits the label HERG for high excitation radio galaxy. Black hole spin must decrease and this is the only parameter that determines the time evolution of the system as seen by the weakening of the two jet production mechanisms (the BZ and BP effects). Figure 4 allows one to understand this model behavior in greater detail. In counterrotating configurations, the inner edge of the disk sits further away from the black hole as seen in the upper panel. As a result of the rapid in-fall onto the black hole in the gap region between the disk and the horizon, the magnetic field is dragged effectively onto the black hole from the inner disk edge. In the co-rotating configuration as seen in the lower panel of Figure 4, instead, the disk inner edge lives closer to the horizon. This ensures that greater magnetic field diffusion occurs in the disk compared to the retrograde configuration creating a weaker black hole-threading magnetic field. In addition to this, the corotating accretion disk has greater efficiency due to the vicinity of the disk inner edge to the black hole. This produces greater disk winds that suppress the jet (Neilsen & Lee 2009; Garofalo, Evans & Sambruna 2010; Ponti et al 2012). As a result of the inherent spin down of the black hole in retrograde configurations, there is a fixed time evolution in the model that cannot be reversed, namely jetted QSOs evolving into non jetted QSOs. Because accretion remains in a radiatively efficient state in the model, it is straightforward to evaluate the time it takes the black hole to evolve to different spin values. Disk efficiency varies from 5 to 6 % of $c^2 dM/dt$ as a result of

the inner disk edge being further out in the retrograde case. In the prograde case, instead, the efficiency varies between 6 and 40 % of $c^2 dM/dt$. Because of the lower efficiency in the retrograde case, counterrotating configurations reach their Eddington limits at higher accretion rates compared to corotating ones for a given black hole mass. While the entire retrograde regime requires less than 10 million years to spin the black hole down from highest spin at the Eddington accretion rate, spinning the black hole back up to highest values in the prograde regime requires an order of magnitude greater time. While spinning the black hole up from 0.1 to 0.2 in the prograde regime requires a few million years at the Eddington accretion rate, the time to spin the black hole up by the same difference (i.e. by $\Delta a$ = 0.1) at higher values of spin requires closer to 10 million years (Kim et al 2016). Once the black hole spin crosses the threshold high prograde value for jet suppression (a feature of the model that remains uncertain), the jet disappears and the object become a radio quiet AGN or radio quiet quasar (Upper panels of Figures 3 and 4).

These radio loud/radio quiet states, their black hole spin values, and the timescales between them are shown in Table 1. With these ideas and timescales it is possible to construct the evolution of a distribution of radio loud and radio quiet quasars from some given initial distribution. Because the rQSOs of Rosario et al (2020) are considered to be merger-induced (Fawcett et al 2020), we will use them as the initial distribution (red curve of Figure 5). In other words, they are merger-induced radio loud and radio quiet QSOs as opposed to QSOs that are in their later stages. It is important to point out that the model (Garofalo, North, Belga & Waddell 2020) and the observations both indicate that the radio quiet population is larger than the radio loud population. Only up to 1 in 5 AGN is radio loud. The fact that the rQSO curve is at least roughly compatible with this fact gives further confidence in the reasonableness of the ideas presented here. It is also important to recognize that the model is founded on prolonged accretion and as such is compatible with the Soltan argument (Soltan 1982; Elvis et al 2002; Garofalo et al 2016). Chaotic accretion is not a feature of the model.

| AGN subclass | Spin/Orientation | Time |
|---|---|---|
| Radio loud or strongly jetted | High/Counterrotation | 0 |
| Radio quiet or weakly jetted | Zero/NA | $8 \times 10^6$ years |
| Radio quiet or weakly jetted | High/Corotation | $10^8$ years |

Table 1: For high spinning black holes born in counterrotating disk configurations that accrete at the Eddington limit, they spin down to zero spin in less than $10^7$ years which makes them jet-less because of the spin value. High spinning corotating, Eddington limited black holes, are also jet-less due to jet suppression, but it takes more than an order of magnitude more time compared to the radio loud counterrotating phase to reach. These are the timescales behind the construction of Figure 5.

In order to produce the light blue curve of Figure 5 we consider the eight data points of the red curve of Figure 2 and associate a spin value to each point based on the model. The far right of Figure 5 corresponds to high retrograde spin (as compatible with Figure 3 and Figure 4). We consider the evolution of each point over a time interval of 10 million years. Over such a timescale no retrograde spinning black hole remains in retrograde configuration for Eddington-limited accretion. This comes from the aforementioned fact that Eddington-limited accretion takes less than 10 million years to spin down to zero spin a rapidly spinning retrograde accreting black hole. Hence, all of the objects on the radio loud side of the dividing dashed line end up on the left hand side. However, mergers occur during that time which replenish the right hand side but no expectation exists that the right hand side be constant. What the model predicts is an rQSO right hand side that is less than the rQSO left hand side and the same for cQSOs. Because of the rapid time evolution from the right hand side of Figure 5 to the left hand side (i.e. from the radio loud phase to the radio quiet one), the right hand side cannot grow. In fact, it decreases a little in the region toward the dividing line. On the left hand side, instead, the time evolution is slower as described above, since a progressively greater amount of mass is needed to spin the black hole up by some $\Delta a$ as the spin increases in the prograde direction. Hence, the objects that were initially on the right hand side of Figure 5 are piling up on the left hand side so the light blue curve increases beyond the red curve. Mergers that produce new radio loud QSOs appear on the right hand side as radio loud rQSOs while the objects that previously occupied the right hand side, begin to contribute to the curve on the left hand side as cQSOs. In short, the total number of objects on the left hand side of Figure 5 increases.

The green curve of Figure 5 represents the objects that made up the red curve of Figure 2 after 100 million years, a timescale that is sufficient for moving all the objects of the red curve to the high prograde regime, which represents radio quiet objects (as illustrated in Figure 3 top panel and Figure 4 bottom panel). Again, this comes from the fact that it takes an accreting black hole about 100 million years to spin up to high prograde spin from an original spin that is very high but in retrograde configuration. Since the evolution at high prograde spin is the slowest (i.e. the time to spin the black hole up by $\Delta a = 0.1$ at high spin is longer than at low spin), the numbers begin to pile up, shifting the curve progressively away from the dashed line while at the same time increasing and shifting the peak to the left. Mergers continue to replenish the radio loud regime but again the rapid time evolution during that phase makes it so the green curve does not increase relative to the other two on the right hand side. It is important to note that one cannot add the blue and green curves of Figure 5 and construct ratios of rQSOs to cQSOs to compare to Figure 1. This is because the blue and green curves are future states of the rQSOs and not the distribution of rQSOs and cQSOs at some moment in time. What can be predicted from Figure 5, however, is that radio quiet cQSOs outnumber all other objects, namely radio loud cQSOs and all rQSOs at any time. This is because at some moment in time the predicted cQSO curve would amount to the sum of previous rQSOs spanning a history of order 100 million years or more.

Despite the fact that Figure 5 involves the same population of objects at separated times while Figure 2 shows different populations at the same time, there are insights to be gained. The light blue and green curves of Figure 5, for example, which represent cQSO curves, have peaks that are higher than the light blue cQSO curve of Figure 2. This is due to the direct connection between the red, light blue, and green curves of Figure 5, whereas Figure 2 describes two separate populations observed at a given time. As mentioned in the previous paragraph, making a direct comparison with Figure 2 for the cQSOs requires reconstructing their history from previous rQSOs. But this would produce a cQSO distribution that is the sum of blue and green like curves that would make the numbers of cQSOs greater than that for rQSOs, at least in the radio quiet regime, but likely also in the radio loud one as well. In fact, the enhancement predicted in the cQSO curve for the radio loud regime requires taking a closer look at Figure 3. Notice the absence of jets for the upper panel. This was described in terms of jet suppression. But the second to top panel shows a prograde system with a jet. Whereas a weak jet of FR0 morphology lives in the $0.1 < a < 0.2$ range of prograde spins (Garofalo & Singh 2019), an FRI jet may still live in radiatively efficient states for disk winds that are not sufficiently strong to generate jet suppression. The values of spin for jet suppression remain uncertain but have been estimated heuristically at $a > 0.7$. When QSOs reach the prograde regime, therefore, there is a period during which a jet of FRI morphology is present that involves some tens of millions of years. Some of these FRI QSOs will have jets whose values are below the observational threshold and will only be observed when their black hole spins have sufficiently increased. However, because the slowest evolution occurs at the highest spins, radio quiet QSOs will live longer for a given Δa (top panel of Figure 3) than these FRI QSOs (second to top panel of Figure 3). A radio quiet QSO, in fact, remains a radio quiet QSO as long as there is fuel because high spinning prograde accreting systems cannot evolve into other spin states. The quantitative details, again, depend on the spin value for jet suppression but qualitatively we can predict an additional subgroup of radio loud cQSOs. In addition to this, there should be post-merger systems whose black hole spins are prograde and roughly in the range $0.2 < a < 0.7$. Such newly born objects would add to the radio loud rQSOs as well as to the radio loud cQSOs. In other words, a newly born prograde accreting QSO with black hole spin of, say, 0.3 would be a radio loud rQSO while it would be a radio loud cQSO when it evolves over some tens of millions of years to a black hole spin of, say, 0.6. These radio loud FRI QSOs have already been explored in the model and have been understood to help explain the dearth of FRI quasars (Kim et al 2016; Garofalo, Joshi et al 2020). In this paper, they help us understand where to look in the model to generate expected total numbers of cQSOs and rQSOs in the radio loud regime. Despite the presence of radio loud prograde accreting QSOs predicted from theory, it should be emphasized that that their jets are not the strongest. As seen in Figure 3, the BZ and BP mechanisms are not maximized (the blue and green arrows are not as large as in the high retrograde case) for the range of spin values that allow prograde jets. In short, the model predicts the existence of more cQSOs across the radio loud to radio quiet regime but that a greater part of these objects will not meet observational thresholds.

It is important to note that at zeroth order the model predicts the behavior shown in Figure 5 with both an increase and a shifting toward the radio quiet region of the peaks over time. The detailed behavior of the red, blue, and green curves in the radio loud regime of Figure 5, however, requires a more in-depth analysis. Those curves will either increase or decrease in number depending on the merger rate and the ensuing accretion rates. Consider, by example, the prediction for two different sets of equal numbers of newly triggered counterrotating black holes whose accretion rates live in the range $0.1(dm/dt)_{Edd} < dm/dt < (dm/dt)_{Edd}$ – where $(dm/dt)_{Edd}$ is the Eddington accretion rate - and in the lower range $0.01(dm/dt)_{Edd} < dm/dt < 0.1(dm/dt)_{Edd}$ as compatible with observations (Figure 15 of Klindt et al 2019). While both sets constitute radiatively efficient accretors, for the objects accreting at lower rates, the evolution is slower which means the blue curve and possibly the green curve will be enhanced in number in the radio loud region. This is because the radio loud phase lasts longer. For the objects accreting at higher rates, the evolution is faster which means the blue curve, and even more so the green curve, will decrease in number in the radio loud region. This is because the radio loud phase lasts less. A similar analysis applies to the radio loud objects in the prograde regime. They must evolve into radio quiet objects but will do so in a time that depends on the rate of accretion. For lower accretion rates, the 0.2 < $a$ < 0.7 range will take longer to pass through, which will add objects to the blue curve and possibly the green curve in the radio loud region. If accretion rates are higher, that range of spin will take less time to pass through, which will add less objects to the blue curve and even less to the green curve in the radio loud region. But while relative differences between the red, blue, and green curves in the radio loud regime can be understood in terms of accretion rates, the evolution on longer timescales will invariably place most of the objects on the left hand side of Figure 5.

Figure 5 is also grounded in the assumption that we are dealing with objects that characteristically tend to dominate in more isolated environments. In richer environments, on the other hand, the model predicts characteristic final prograde states that are not radio quiet AGN or radio quiet quasars (e.g. Garofalo, 2019). In Garofalo 2019, for example, one finds a greater fraction of radio loud quasar phases associated with both a transition toward the prograde spin state but also with a transition in the state of accretion. Instead of remaining in cold mode accretion states like the objects of this paper, such systems evolve away from radiatively efficient accretion and into advection dominated accretion modes. Such objects tend to have larger black holes, stronger jet feedback, and are classified as low excitation radio galaxies, i.e. not as cQSOs. Hence, another prediction of the model is that the objects described in this paper should tend to distribute themselves more in environments with smaller dark matter halos and smaller black holes. While this would extend our work beyond the confines of this paper, the model actually predicts that the peak of the cQSO curve in the radio quiet regime for objects that inhabit richer environments, is lower than the peak for cQSOs in the radio quiet regime associated with more isolated environments. It would be interesting to explore the cluster richness of the objects in Figure 2 to test this idea. It is clear, however, that the numbers of cQSOs are predicted to be larger than the rQSOs as is observed (Richards et al 2003; Klindt et al 2019).

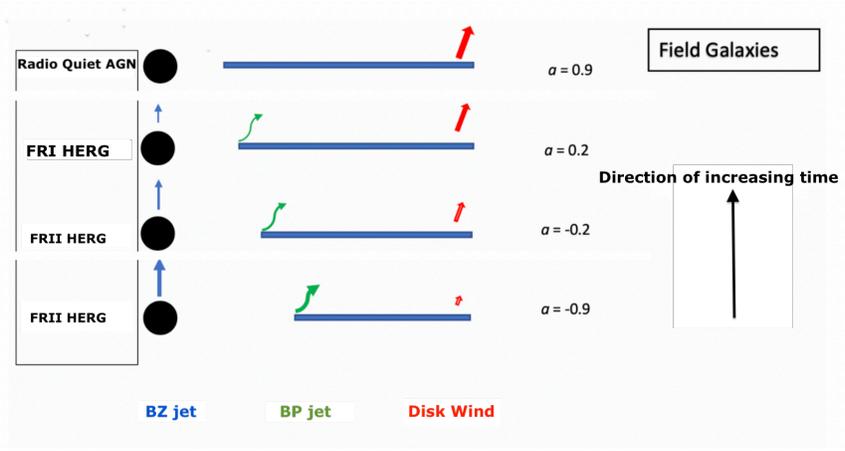

Figure 3: Time evolution of radio loud quasars for accreting black holes with relatively weak feedback. Leftmost column describes jet morphology and excitation state; BZ represents strength of the Blandford-Znajek jet; BP stands for Blandford-Payne jet; Disk wind indicates the strength of the disk outflow; spin indicated on the far right with negative values corresponding to retrograde accretion and positive values prograde accretion. The weak feedback is due to smaller black holes that are formed in more isolated environments. Hence the classification "Field Galaxies" in the upper right.

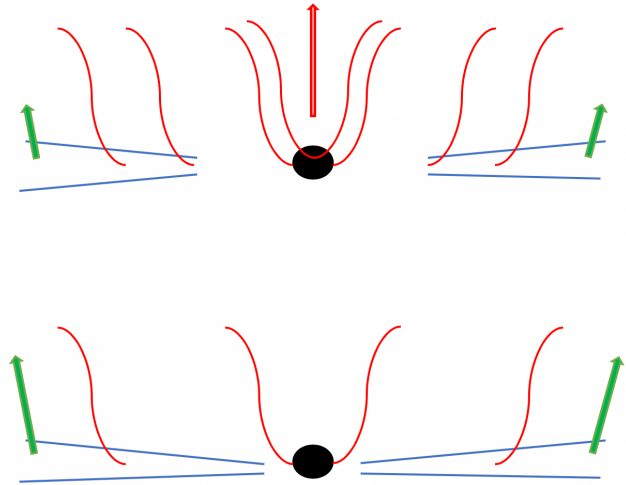

Figure 4: Retrograde accreting black hole (top panel) and prograde accreting black hole (lower panel). Because of the large gap region for counterrotation, the magnetic flux on the black hole is enhanced and a strong jet (red arrow) is produced. Because of the small gap region for co-rotation, the magnetic flux on the black hole is weaker, but more importantly the disk wind (green arrow) is strong enough to suppress jet formation regardless. The bottom panel represents a bright radio quiet QSO while the upper panel represents a radio loud QSO.

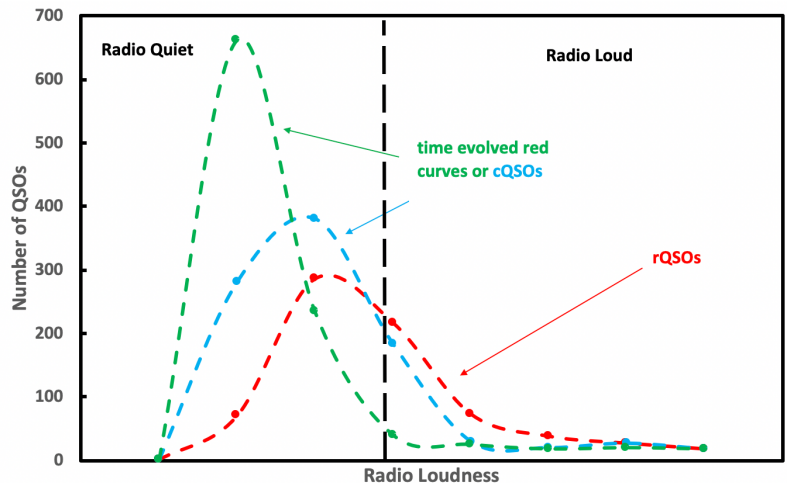

Figure 5: rQSOs from Rosario et al (2020) subjected to the time evolution in the model of Figure 3. Blue curve is the red curve after 10 million years while the green curve is the red curve after 100 million years. Mergers leading to radio loud rQSOs that evolve slowly feed the blue and green curves at the far right end.

3. Conclusions

We have shown that observations of red and blue quasars support a time evolution framework in which radio loud quasars evolve into radio quiet quasars. The only tool needed to reproduce our results is the time to spin black holes down and up to different values at or near the Eddington accretion rate, which models radiatively efficient thin disks. Because retrograde accretion is unstable and mergers are expected to produce more prograde accreting black holes, red QSO curves should have greater areas under the curve on the radio quiet side. Because our model predicts that radio loud quasars evolve rapidly into radio quiet quasars, and that radio quiet quasars linger in such states unlike their radio loud ancestors, the number of radio quiet quasars must be greater than the radio loud quasars. This is compatible with the cQSO curves modeled in this paper. Our time evolution picture makes sense of additional puzzling features of the observations of Rosario et al (2020): rQSO and cQSO curves that approach each other at both extremes of the radio loud/radio quiet spectrum but with higher numbers on the radio quiet side; and, a noticeable difference in the region near the boundary line at low radio loudness, where more compact jets are observed. These compact jets are not young or emerging AGN in the model, but objects that have experienced a rapid retrograde accreting phase and are now approaching zero spin in their black holes, so their jets are turning off. It is interesting to point out that near this boundary line are the objects that constitute X-shaped radio galaxies (Garofalo, Joshi et al 2020).

Acknowledgments

It is a pleasure to recognize the unusual depth of our referee who not only understood our model but was always a step ahead of us, pointing out issues that needed resolution and/or clarification.